\documentclass[aoas]{imsart}

\RequirePackage{amsthm,amsmath,amsfonts,amssymb}
\RequirePackage[authoryear]{natbib}
\RequirePackage{graphicx}
\usepackage{subcaption, multirow,hyperref} 
\startlocaldefs

\endlocaldefs

\begin{document}

\begin{frontmatter}
\title{Derivation of Dietary Patterns dependent on Diabetes status using ordinal Supervised Robust Profile Clustering: Results from Hispanic Community Health Study/Study of Latinos }
\runtitle{Ordinal Supervised Robust Profile Clustering}

\begin{aug}
\author[A]{\fnms{Briana Joy}~\snm{K. Stephenson}\ead[label=e1]{bstephenson@hsph.harvard.edu}\orcid{0000-0002-6147-1039}},
\author[B]{\fnms{Daniela}~\snm{Sotres-Alvarez}\ead[label=e2]{dsotres@email.unc.edu}\orcid{0000-0002-3226-6140}}, 
\author[C]{\fnms{Martha}~\snm{Daviglus}\ead[label=e3]{daviglus@uic.edu}\orcid{ 0000-0002-6791-8727}}, 
\author[D]{\fnms{Ramon A.}~\snm{Durazo-Arvizu}\ead[label=e4]{rdurazoarvizu@chla.usc.edu}\orcid{0000-0001-6563-7107}}, 
\author[E]{\fnms{Yasmin}~\snm{Mossavar-Rahmani}\ead[label=e5]{yasmin.mossavar-rahmani@einsteinmed.edu}\orcid{0000-0002-9214-6124}}, 
\and
\author[B]{\fnms{Jianwen}~\snm{Cai}\ead[label=e6]{cai@bios.unc.edu}}
\address[A]{Department of Biostatistics, Harvard T.H. Chan School of Public Health\printead[presep={,\ }]{e1}}

\address[B]{Department of Biostatistics, University of North Carolina at Chapel Hill\printead[presep={,\ }]{e2,e6}}
\address[C]{Department of Medicine, University of Illinois at Chicago College of Medicine\printead[presep={,\ }]{e3}}
\address[D]{Division of Research on Children, Keck School of Medicine at University of Southern California \printead[presep={,\ }]{e4}}
\address[E]{Department of Epidemiology and Population Health, Albert Einstein College of Medicine\printead[presep={,\ }]{e5}}

\end{aug}

\begin{abstract} 
The burden of diabetes has disproportionately impacted Hispanic/Latino residents in the United States, with diet recognized as a major modifiable risk factor. Outcome-dependent dietary patterns provide insight into what foods may be associated with the increased severity and progression of diabetes. However, the ethnic and geographical heterogeneity of US Hispanic/Latino adults makes it difficult to identify and distinguish differences within their diet as risk increases. Supervised robust profile clustering (sRPC) is a flexible joint model that can identify dietary patterns associated with diabetes, while partitioning out those defined by their ethnicity and geography. However, sRPC has only been applied to binary outcomes. We extend the existing model to develop the ordinal sRPC.  Using baseline dietary data (2008-2011) from the Hispanic Community Health Study/Study of Latinos, we illustrate the utility of our model to identify dietary patterns associated with the three-levels of diabetes status (i.e. normal, pre-diabetes, diabetes). Simulation studies confirmed that ordinal sRPC improved identification and characterization of these patterns compared to a standard supervised latent class model. Results indicated that participants who had greater consumption of fruits, snack foods, and refined grain breads may be more likely to be associated with an increasing severity of diabetes status. 
\end{abstract}


\end{frontmatter}


\section{Introduction}

Nutrition epidemiology is centered on the understanding of diet-disease relationships. The consumption of foods is highly correlated, and seldom consumed in isolation, warranting the analysis of dietary patterns to better examine the diet disease relationship \citep{hu2000prospective,pate2015associations}. Dietary patterns can be derived a priori, where patterns are predefined and the study population is evaluated based on its adherence to the referenced pattern. They can also be a posteriori where the patterns are reflective of the habits and behaviors prevalent in the study population. For the scope of this paper, we will focus on a posteriori patterns, which tend to be more data driven and independent of researcher-defined priorities. The analysis of diet-disease relationships via a posteriori dietary patterns fall under two techniques: unsupervised and supervised. In an unsupervised setting, the dietary pattern is analyzed first, typically via a factor or cluster-based analysis, and subsequently treated as an exposure in an outcome analysis, such as a regression or chi-squared test \citep{maldonado2022posteriori,peters2022western,owczarek2022nutrient,zheng2017relationship,sotres2013maternal}.  In a supervised setting, dietary patterns are derived and dependent on predefined known outcome. A commonly used approach in nutrition epidemiology is the reduced rank regression, where linear combinations of different food components are treated as factors in a regression model, where they are used to explain the variation in an outcome \citep{hoffmann2004application}. The outcome is typically a surrogate or intermediary variable (e.g. biomarker) that is able to provide a physiological measure for a disease of interest \citep{tucker2010dietary,schulze2018food}.  

An often noted limitation of dietary pattern analysis is the dependency on the analytic sample from the study population. Large heterogeneous populations are often prone to be overpowered by majority subgroup behaviors. This leaves smaller-sized subgroups less identifiable if they do not generalize with patterns fit with the larger majority. Some studies have avoided this by removing the smaller-sized subgroups that may introduce noise to the overall patterns and focusing the attention on that larger majority. This global clustering assumption becomes impractical in highly diverse populations. Subjects may have different consumption behaviors of specific food items based on their background or other life factors. 

Bayesian techniques allowed more flexible models for pattern identification in the presence of subgroup heterogeneity. Multi-study factor analysis and robust profile clustering are recent approaches that focus on subpopulation heterogeneity with applications in nutrition epidemiology \citep{de2019shared,de2022shared,stephenson2023racial,stephenson2020empirically}. Multi-study factor analysis (MSFA) is an extension of exploratory factor analysis, where subsets of highly correlated variables are clustered to explain the variation in the overall population, as well as variable subsets explained within known subpopulations \citep{de2019multi}. Robust profile clustering (RPC) shares a similar dual flexibility with MSFA with additional flexibility in data type flexibility (e.g. continuous, non-normal, binary, categorical). Extended from a standard finite mixture model, RPC clusters a subset of exposure variables and subjects that share behaviors across the entire population, and isolates those that deviate from those overall patterns into a localized clustering pattern based on the subjects' known subpopulation or demographic \citep{stephenson2019robust}. 

While MSFA is tailored for an unsupervised approach, RPC has been extended to a supervised setting, to allow the derived global exposure patterns to cluster in accordance with their dependence on a known binary outcome \citep{stephenson2022derivation}. This is achieved through a joint model that allows for the global cluster behaviors and probability of the binary outcome to borrow information from one another to better inform outcome-dependent exposure patterns and parameter estimation. 

Hispanic/Latino residents have experienced a growing burden of diabetes in the United States \citep{umpierrez2007diabetes, aviles2017sea}. Yet, heterogeneity within the Hispanic/Latino community indicates that the burden of diabetes is not distributed equally across different ethnic backgrounds \citep{schneiderman2014prevalence,heiss2014prevalence,maldonado2022posteriori}. Diet is a modifiable risk factor of diabetes status that can vary significantly within this community by both cultural and geographic differences \citep{maldonado2021dietary,stephenson2020empirically,de2022shared}. With a focus on the progressive risk level of diabetes in Hispanic/Latino adults and the known ethnic and geographic heterogeneity of dietary intake, we extend the supervised robust profile clustering to accommodate an ordinal outcome, in order to better identify population-based dietary patterns associated with normal, pre-diabetic, and diabetic levels of Hispanic/Latino adults in the United States. 

We organize the paper as follows: Section 2 introduces the ordinal supervised robust profile clustering model. Section 3 provides a small simulation study to highlight the properties of the supervised RPC compared to other supervised methods. Section 4 applies the model to the Hispanic Community Health Study/Study of Latinos. Section 5 provides a brief discussion of the model and details future directions.  

\section{Ordinal Supervised Robust Profile Clustering Model}
\subsection{Supervised Robust Profile Clustering}
The supervised robust profile clustering (supRPC) model was introduced by \citet{stephenson2022derivation} to provide a dimension reduction technique that captured shared patterns amongst a general population that are associated with an outcome, while simultaneously removing subpopulation-specific differences which can obfuscate relationships. The model is the supervised counterpart to the unsupervised model of the same name \citep{stephenson2019robust}. Observed data includes a set of $p$ exposures $\mathbf{x}_{i\cdot}=(x_{i1}, \ldots, x_{ip})$, a binary outcome $y_i \in (0,1)$, a set of observed $g$  covariates that should be adjusted for when quantifying the association to the binary outcome, $\mathbf{w}_i =(w_{i1},\ldots, w_{ig})$ and a subpopulation identifier $s_i \in (1,\ldots, S)$.  The model has two parts. The first part uses the observed exposure data to generate a flexible mixture model that clusters shared behaviors of subjects of the overall population in a global cluster and shared behaviors of subjects within a known subpopulation in a localized cluster. The second part uses the observed binary outcome and derived global cluster assignment from part 1 to generate a probit regression model to quantify the association of the global cluster membership on the probability of a positive outcome. As a joint model estimated under a Bayesian framework, the global cluster assignment and binary outcome are informed from each other. In other words, the probability of being assigned to a global cluster is dependent on both the shared behaviors within that global cluster as well as the proportion of subjects in that cluster that share that outcome. 

We introduce the parameters of the supRPC model below, where $K_0$ and $K_s$ denote the finite number of global and subpopulation-specific local clusters in the model. Let $C_i$, denote membership to a global cluster, such that $Pr(C_i= h) =\pi_h$ and $\sum_{h=1}^{K_0}=\pi_h = 1$. Let $L_{ij}$, denote membership to a local cluster, such that $Pr(L_{ij}= l|s_i= s) = \lambda_{l=1}^{(s)} and \sum_l^{K_s}=\lambda_l^{(s)}=1$. Here, individuals within the same subpopulation assume their own localized clustering pattern. Given the partition of items at global or local levels, $G_{ij}$, denotes a binary indicator for each item $j\in (1,\ldots,p)$ and individual $i\in (1,\ldots,n)$, determining whether that item is reflective of the global pattern assigned $(G_{ij}  =1)$ or local pattern assigned, $(G_{ij}=0)$, such that $Pr(G_{ij}=1) = \nu_j^s$. The patterns reflected at the global and local level are described through the set $\Theta_{0\cdot\cdot}=\{\theta_{0C_i} \}_{j=1}^p$ and $\Theta_{1\cdot\cdot}^{(s)}=\{\theta_{1jL_{ij}}^{(s)}\}_{j=1}^p$, respectively. Participants who observe consumption of food item $j$ at level $r$ are defined by the identity function, $1(x_{ij}=r)$. Each individual variable $\theta_{\cdot k}$ is a multivariate categorical variable with $d_j$ levels. The set of covariates adjusted for association with the outcome, including a covariate indicating an individual’s global cluster assignment is described with $\xi$ in the probit regression model. Using $\Omega$ to denote the full set of parameters, we describe the likelihood as follows, 

\begin{equation}
\mathcal{L}(x_i, y_i, s_i, w_i|\Omega) =\\
 \left[ \sum_{k=1}^{K_0} \pi_k \prod_{j:G_{ij}=1}^p \prod_{r=1}^{d_j} \theta_{0jk,r}^{1(x_{ij}=r)} \Phi(w_i \xi)^{y_i}[1-\Phi(w_i\xi)]^{1-y_i}\right] \prod_{j:G_{ij}=0}^p \sum_{l=1}^{K_s} \lambda_l^{s_i} \prod_{r=1}^{d_j} \theta_{1jl,r}^{s_i, 1(x_{ij}=r)}
\end{equation}

\subsection{Ordinal Supervised Robust Profile Clustering (osRPC)}
The ordinal supervised robust profile clustering model extends the binary outcome to an ordinal outcome, allowing $M$ different ordered levels, $y_i \in (1,2, \ldots, M)$.
 
\begin{equation}
L(x_i, y_i, s_i, w_i|\Omega) = \\
\left[ \sum_{k=1}^{K_0} \pi_k \prod_{j: G_{ij}=1}^p \prod_{r=1}^{d_j} \theta_{0jk,r}^{1(x_{ij}=r)} \prod_{m=1}^M Pr(y_i=m|w_i, \xi)^{1(y_i=m)}\right] \prod_{j:G_{ij}=0}^p \sum_{l=1}^{K_s} \lambda_l^{s_i} \prod_{r=1}^{d_j} \theta_{1jl,r}^{1(x_{ij}=r)}
\end{equation}

Under the ordinal probit regression model, we introduce a latent variable $Z_i$ to transform the ordinal probability outcomes on the normal real space with $M$ boundary points $\{\gamma_0 = -\infty < \gamma_1 < \gamma_2 < \ldots < \gamma_M=\infty\}$, where $Pr(y_i = m|\xi, \gamma_m, \gamma_{m-1})$ is bounded between $[\Phi(\gamma_{m-1}-W_i\xi), \Phi(\gamma_m - W_i\xi)]$. 

Integration of the multinomial probit model does not have a closed form. This makes it an ideal candidate for  Bayesian estimation, allowing both structural flexibility and adaptability to the input data. This is achieved by defining distributions of each of the parameters and using a prior distribution to incorporate any prior information we may have on the data \textit{a priori}. Probability parameters $\pi_\cdot, \lambda_l^{(s_i)}, \theta_{0jk,r}, \theta_{1lj,r}^{(s_i)}$, describe categorical variables and assume a multinomial distribution. Exploiting conjugacy, we implement a flat and symmetric Dirichlet prior for each of these parameters. The number of global and local clusters is not known \textit{a priori}. To remedy this uncertainty, the model is overfitted with a large number of clusters to mimic that of an infinite mixture model \citep{van2015overfitting}. The covariate vector, $\xi$, is drawn from a multivariate normal distribution of $q$-length, where $q=g+K_0$ including the number of demographic confounders plus the number of global clusters, indicated by the global cluster membership assignment $(C_i)$. The latent variable $Z_i$ is drawn from a truncated normal based on the bounds respective to the observed outcome $y_i$, with mean $W_i \xi$ and standard deviation 1. The breakpoints separating the ordinal outcomes are updated from a truncated Beta distribution as detailed in \citep{sha2019bayes}. We detail the full posterior computation and inference of all parameters below.

\subsubsection{Posterior Computation of osRPC algorithm}\label{sec:mcmc}
\sdescription{The following steps outline the posterior distributions needed to implement a Gibbs sampling algorithm for osRPC}
\begin{enumerate}
	\item Update the global component indicators $(G_{ij}|s_i=s) \sim Bern(p_{ij})$, for each $i \in (1,\ldots, n)$ with respective subpopulation index $s$, where 
	$$
	p_{ij} = \frac{\nu_j^{(s)}\prod_{r=1}^{d_j} \theta_{0j{C_i},r}^{1(x_{ij}=r)}}{\nu_j^{(s)}\prod_{r=1}^{d_j}\theta_{0j{C_i},r}^{1(x_{ij}=r)} + (1-\nu_j^{(s)})\prod_{r=1}^{d_j}(\theta_{1j{L_{ij}},r}^{(s)})^{1(x_{ij}=r)}}
	$$

	\item Update global cluster index $C_i, i =1, \ldots, n$ from its multinomial distribution where 
	$$
	Pr(C_i=h) = \frac{\pi_h \prod_j \prod_{r=1}^{d_j} \theta_{0jh,r}^{1(x_{ij}=r,G_{ij}=1)} \prod_m \Phi(\gamma_m-W_i\xi)^{1(y_i=m)}}{\sum_k \pi_k \prod_j \prod_{r=1}^{d_j}\theta_{0jk,r}^{1(x_{ij}=r,G_{ij}=1)}\prod_m \Phi(\gamma_m - W_i\xi)^{1(y_i=m)}}
	$$

	\item Update global cluster probability weights, where each global cluster is weighted by the number of participants assigned to that respective cluster for that iteration.
		 $$\pi=(\pi_1,\ldots,\pi_K) \sim \text{Dir}\left(\alpha+\sum_{i=1}^n \mathbf{1}(C_i=1),\ldots, \alpha+\sum_{i=1}^n \mathbf{1}(C_i=K)\right).$$

	\item Update local cluster probability weights for each subpopulation $s \in (1, \ldots, S)$, where each local cluster is weighted by the number of participants assigned to that respective subpopulation-specific cluster for that iteration.
	$$\lambda^{(s)}=\left(\lambda_1^{(s)},\ldots,\lambda_K^{(s)}\right) \sim \text{Dir}\left(\alpha + \sum_{i:s_i=s} \sum_{j=1}^p \mathbf{1}(L_{ij}=1), \ldots, \alpha + \sum_{i:s_i=s} \sum_{j=1}^p \mathbf{1}(L_{ij}=K)\right).$$
	
	\item Update the multinomial parameters, where $\sum_{i:G_{ij}=1,C_i=h} \mathbf{1}(x_{ij}=r)$ is the total number of subjects allocated to global cluster $h$ for variable $j$ with an observed consumption level of $r$, letting $\eta=1$. Similarly, $\sum_{i:G_{ij}=0,L_{ij}=h,s_i=s} \mathbf{1}(x_{ij}=r)$ is the total number of subjects allocated to local cluster $h$ within subpopulation $s$ for variable $j$ and an observed consumption level of $r$.
\begin{align*}
\theta_{0jh,\cdot} &\sim \text{Dir}\left(\eta+\sum_{i:G_{ij}=1,C_i=h} \mathbf{1}(x_{ij}=1),\ldots,\eta+\sum_{i:G_{ij}=1,C_i=h} \mathbf{1}(x_{ij}=d)\right) \\
\theta^{(s)}_{1jh,\cdot} &\sim \text{Dir}\left(\eta+\sum_{i:G_{ij}=0,L_{ij}=h,s_i=s} \mathbf{1}(x_{ij}=1),\ldots,\eta+\sum_{i:G_{ij}=0,L_{ij}=h,s_i=s} \mathbf{1}(x_{ij}=d)\right)
\end{align*}

\item Update $\nu_j^{(s)} \sim \text{Be}(1+\sum_{i:s_i=s} G_{ij}, \beta^{(s)} + \sum_{i:s_i=s} (1-G_{ij}))$.

\item Update Beta-Bernoulli hyperparameter: $\beta^{(s)} \sim \text{Ga}(a_{\beta}+p, b_{\beta} - \sum_{j=1}^p \log(1-\nu_j^{(s)}))$.

\item Draw subject-specific latent response variable for probit model for a 3-level outcome, $z_i$ for $i \in (1,\ldots,n)$:
$$
z_i \sim \begin{cases}
	N_{z_i \in (-\infty,\delta_1)}(W\xi,1) &y_i=1 \\
	N_{z_i \in (\delta_1, \delta_2)}(W\xi,1) & y_i=2 \\
	N_{z_i \in (\delta_2, \infty)}(W\xi,1) & y_i=3
	\end{cases}
$$

\item Update regression coefficients: $\xi \sim MVN((\Sigma_0^{-1} + W'W)^{-1}(\Sigma_0^{-1}\xi_0 + W'Z) ,(\Sigma_0^{-1} + W'W)^{-1})$, where $\xi_0 \sim MVN(\mu_0, \Sigma_0)$.

\item Update augmentation variable $\delta_1$ for ordered probit, where $c_{11}=\max_i{(z_i|y_i=1)} $and $c_{12}=\min_i{(z_i|y_i=2)}$.
$$
\begin{aligned}
\delta_1^* &\sim U\left(\frac{\Phi(c_{11}/s_0)}{\Phi(\delta_2/s_0)},\Phi(c_{12}/s_0)\right) \\
\delta_1 &= \Phi^{-1}(\Phi(\delta_2/s_0)\delta_1^*)
\end{aligned}
$$

\item Update augmentation variable $\delta_2$ for ordered probit, where $c_{21}=\max_i{(z_i|y_i=2)} $and $c_{22}=\min_i{(z_i|y_i=3)}$.
$$
\begin{aligned}
\delta_2^* &\sim U\left(\frac{\Phi(c_{21}/s_0)-\Phi(\delta_1/s_0)}{1-\Phi(\delta_1/s_0)}, \frac{\Phi(c_{22}/s0)}{1-\Phi(\delta_1/s_0)}\right) \\
\delta_2 &= \Phi^{-1}\left(\Phi(\delta_1/s_0)+\delta_2^*[1-\Phi(\delta_1/s_0)]\right)
\end{aligned}
$$
\item Calculate probit likelihood for all individuals, 
$$\Phi(W_i\xi)^{1(y_i=1)}\times \left[\Phi\left(\frac{\delta_2-W_i\xi)}{s_0}\right) - \Phi\left(\frac{\delta_1-W_i\xi}{s_0}\right)\right]^{1(y_i=2)}\times\left[1-\Phi\left(\frac{\delta_2-W_i\xi}{s_0}\right)\right]^{1(y_i=3)}
$$

\end{enumerate}

MCMC estimation was performed in MATLAB 2022 (v2022b). An adaptive sampler of 10,000 runs with a burn-in of 5,000 to determine the number of global and local clusters, where $K_0$ and $K_s$ were both initialized at 50, consistent with an overfitted mixture model framework. A cluster was defined as nonempty if at least 5\% of participants were assigned to it. The derived nonempty global and local clusters were then implemented in a fixed sampler of 25,000 runs, burn-in after 15,000 and thinning every 10 iterations. Dietary patterns were defined via the modal consumption level of each individual food distribution within that global or local cluster. Label switching was imposed every ten iterations using the random permutation sampler to encourage mixing \citep{fruhwirth2001markov}. Non-empty global and local clusters were preserved for post-processing after MCMC was complete. Labels were reordered and relabeled, also in post-processing, using a similarity matrix from the MCMC output of participant assignments and a hierarchical clustering with complete linkage approach as described in \cite{stephenson2019robust, medvedovic2002bayesian,stephenson2022derivation}. Posterior median estimates were taken for all model parameters and used to assign participants to a respective global or local cluster using the highest posterior median probability of assignment.

\section{Simulation}
We tested the model performance of the ordinal supervised robust profile clustering (osRPC) model against a similar joint model of a latent class model with an ordinal probit regression, referred here as an ordinal supervised latent class model (osLCM) under two settings \citep{desantis2012supervised,larsen2004joint}. Case A: global only, compared the two models when no subpopulation-specific deviations were expected. Case B: global-local hybrid, compared the two models when subpopulation-specific deviations were present allowing for global and local heterogeneity. 

A total of 500 simulated datasets were generated and compared across the case studies. Each simulated dataset contained 4 subpopulations of equal size $(n_s=1200)$. Three distinct global patterns were generated with $p=50$ variables containing $d_j=d=4$ response levels. Individuals were pre-assigned to one of the three global patterns with equal probability in both cases A and B. Expected response levels for a given global pattern were favored with a 0.85 probability, and 0.05 probability for all other possible responses. Global pattern 1 favored a response level of 3 for the first 25 variables and a response level of 1 for the remaining variables. Global pattern 2 favored a response level of 2 for the first ten variables, and a response level of 5 for the remaining forty variables. Global pattern 3 favored a response level of 1 for the first ten variables and a response level of 2 for the subsequent 20 variables, followed by a response level of 3 for the remaining 20 variables. Probability of a positive outcome varied based on global pattern assignment. Individuals assigned to global pattern 1, were identified as high risk and had outcome response probabilities ranging from 0.97-0.99, with probabilities showing a small variation by subpopulation.  Individuals assigned to global pattern 2, were identified as low risk and had outcome response probabilities ranging from 0.14-0.22, with variation by subpopulation. Individuals assigned to global pattern 3, were identified as medium risk with outcome response probabilities ranging from 0.46-0.58, with variation by subpopulation. Both models were overfit with $K_0=K_s=50$ clusters. Adaptive and fixed samplers were run under the same MCMC settings previously described in \ref{sec:mcmc}. Model fitness was measured using the deviance information criterion (DIC) as defined by \cite{celeux2006deviance}. Model estimates of the probability of outcome, $Y_i$ and global food item indicator, $\nu$ were compared to true estimates using mean squared error (MSE). 
 
\subsection{Results} 
Both models showed good mixing and convergence under both simulation cases. Table \ref{sim-t} provides a summary of results. Under case A (global only setting), both methods performed relatively well. The osLCM had a slightly better fit (lower DIC) to the data as it provided more parsimonious parameterization. The additional parameters that comprise the local level clustering in the osRPC added an unnecessary complexity as no local clustering was identified. However, the osRPC was able to identify that a global-only clustering model was preferred with strong probabilities favoring a global cluster setting for all variables $(\nu_{MSE}=0.03)$. Both models accurately identified the true cluster patterns and ordinal outcome probabilities.  

Under case B (global and local clustering), differences became more evident between the two methods. Pattern identification worsened under the osLCM model. The presence of the local clustering generated more global clusters $(K_{osLCM}=6)$ and lower identifiability of the true patterns. The osRPC model was able to identify both the global and local true cluster pattern, as well as which variables deviated from the global clustering and assumed a local cluster pattern. Even though the patterns were misspecified in osLCM model, the true underlying probability of ordinal outcome was still maintained.  With the added complexity of local clustering present, the osRPC performed better with the data, as full parameterization was utilized for this setting and closely approximated the true estimate values. 

\begin{table}
\caption{Results from simulation cases A (global clusters only) and B (global and local clusters present)}
\label{sim-t}
\begin{tabular}{@{}lcccc@{}}
\hline
 & \multicolumn{2}{c}{Case A} &\multicolumn{2}{c}{Case B} \\
\cline{2-5}
Metric & osLCM & osRPC & osLCM & osRPC \\
\hline
$K_{pred}$ & 3& 3& 6 & 3 \\
\hline
DIC & $3.40\times 10^5$ & $4.61\times 10^5$ & $4.79\times 10^5$ & $4.61\times10^5$ \\
 \hline
 $P(Y)_{MSE}$ & 0.0078 & 0.0074 & $0.0098$ & $0.0074$ \\
 \hline
 $Y_{ord}$ & 0 & $0$ & 0.032& 0 \\
 \hline 
 Pattern 1 Classification & 1 & 1 & 0.75 & 1 \\
 Pattern 2 Classification & 1 & 1 & 0.74 & 1 \\
 Pattern 3 Classification & 1 & 1 & 0.75 & 1 \\
 \hline
 $\nu_{MSE}$ & $-$ & 0.03 & $-$ & 0.01 \\
 \hline
\end{tabular}
\end{table}

\section{Application to HCHS/SOL Data}
\subsection{Description of Data}
The Hispanic Community Health Survey/Study of Latinos is a multi-site community-based cohort study focused on the health and incidence of disease among Hispanic/Latino adults aged 18-74 years residing in the United States. The study enrolled 16,415 participants from 2008-2011 identifying as one of six ethnic backgrounds (Central American, Cuban, Dominican, Mexican, Puerto Rican, South American) across four study sites (Bronx NY, Chicago IL, Miami FL, San Diego CA) under a complex sampling design. Details of the recruitment, selection, and sampling design are described elsewhere \citep{Lavange:2010aa,Sorlie:2010aa}. 

Previous research has indicated that dietary patterns differ by both ethnicity and geography in this cohort \citep{stephenson2020empirically,de2022shared}. For that reason, we have defined our subpopulations as a cross-section of ethnic background and study site. Diet was assessed via a food propensity questionnaire (FPQ). This assessment was developed based on data from the 24-hour recall collected, and curated to include foods and beverages typically seen in a Hispanic/Latino diet. The assessment was collected 12 months after enrollment and includes frequency of consumption data of 132 foods and beverages since enrollment (last 12 months). Three food items were collapsed due to similarities (e.g. white bread or rolls alone was combined with white bread or rolls in a sandwich). Frequency consumption levels ranged from ``never'' to ``six or more times per day''. We have collapsed these frequency levels to five: no consumption, less than two to three times per month, three to four times per week, five to seven times per week, and more than once a day. For ease of interpretability, we will refer to these consumption ranges as never, monthly, weekly, daily, and daily-plus. Further, several foods had small percentages of consumption at certain levels and were collapsed to two, three, or four levels as needed. Distribution of collapsed food items are provided in supplementary based on the sample included for analysis. 

Diabetes status was used as a three-level outcome variable (diabetic, pre-diabetic, non-diabetic). Based on the American Diabetes Association definition \citep{american2010diagnosis}. Diabetic status was defined as participants with one of the following: (1) a fasting time greater than 8 hours and a fasting glucose greater than or equal to 126 mg/dL; (2) a fasting time less than 8 hours and a fasting glucose greater than or equal to 200 mg/dL; (3) a post oral glucose tolerance test greater than or equal to 200 mg/dL; (4) elevated glycated hemoglobin level (A1C $\ge$ 6.5\%); or (5) indication of anti-diabetic medication use. Pre-diabetic status was defined as participants with one of the following: (1) fasting time greater than 8 hours and a fasting glucose in range 100-125 mg/dL; (2) post oral glucose tolerance test in the range 140-199 mg/dL; (3) glycated hemoglobin levels ranging from 5.7\%-6.5\%. Non-diabetic status were participants that did not meet any of the above criteria for diabetes or pre-diabetes. 

All participants who contained reliable responses to the food propensity questionnaire,  complete diabetes status information, and belonged to a subpopulation at least 100 participants in size were included for analysis (n=11,854). A total of 13 subpopulations were included for analysis and are detailed in Table \ref{sub-t}. 
\begin{table}
\caption{Frequency distribution of subpopulations included for analysis. Cell entries denote number of participants included for analysis from that study site and ethnicity cross-section.}\label{sub-t}
	\begin{tabular}{@{}lcccccc@{}}
	\hline
	Study Site/Ethnicity  & C. American & Cuban & Dominican & Mexican & Puerto Rican & S. American \\
	\hline
	Bronx & 165 & NA & 1018 & 145 & 1325 & 133\\
	Chicago & 299 & NA & NA & 1771 & 594 & 253 \\
	Miami & 831 & 1928 & NA & NA & NA & 401 \\
	San Diego & NA & NA & NA & 2991 & NA & NA
	\end{tabular}
\end{table}

\subsection{Outcome dependent global dietary patterns}
While a complex sampling design was utilized in the collection of HCHS/SOL, the scope of this paper focused on characteristics related to the cohort. We present demographic information about the participants included for analysis in Supplementary Table 1. A majority of participants identified as female (61.3\%), unemployed (39.5\%), never used alcohol (53.5\%), income less than \$30,000 (64.8\%), above the age of 45 (61.7\%), and of Mexican background living in San Diego (25.5\%). 

\begin{figure}
\includegraphics[height=0.9\textheight,width=0.8\textwidth]{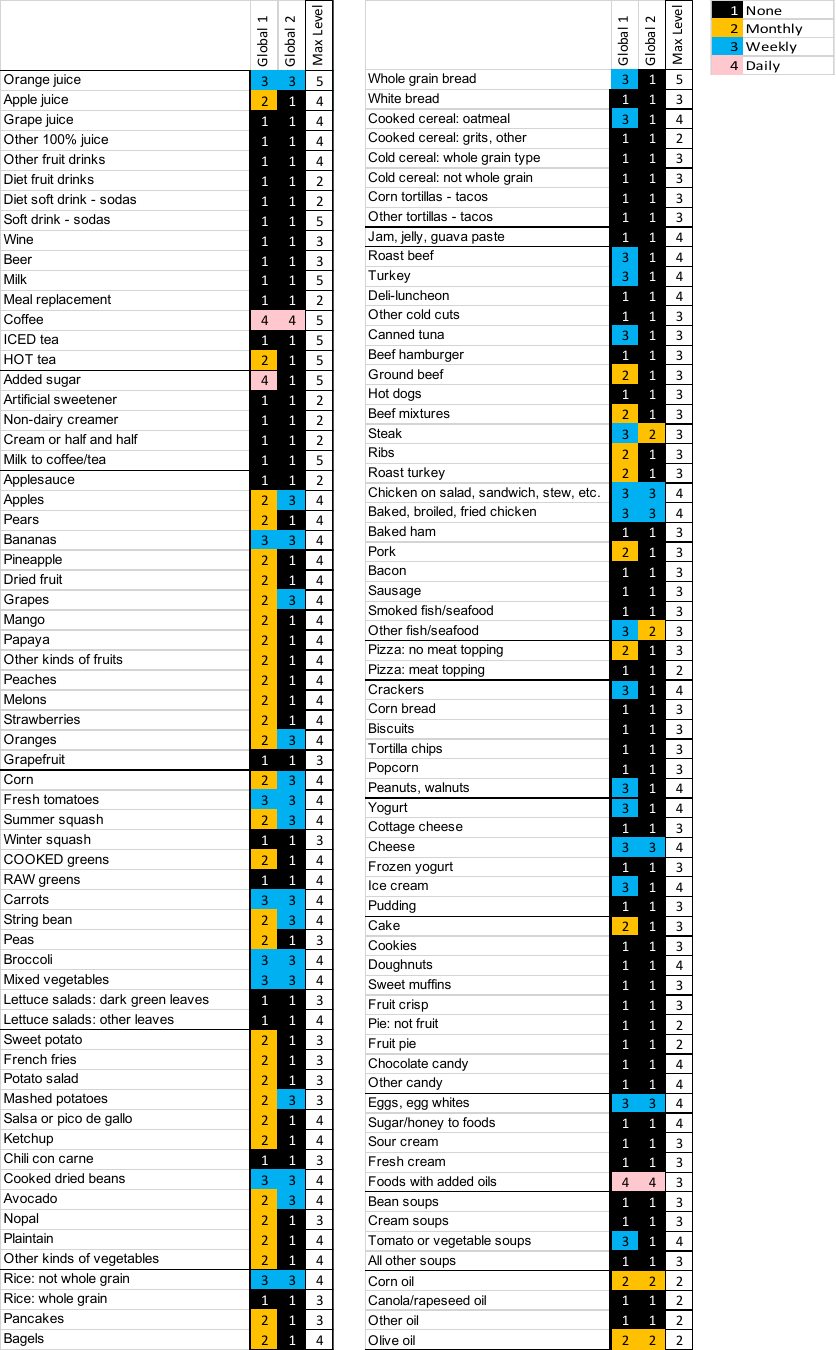}
\caption{Dietary patterns derived at the global level, defined by the  consumption level with highest posterior probability for that food variable. Levels are distinguished by colors and numbers  1/Black=no consumption, 2/Yellow=monthly consumption, 3/Blue=weekly consumption, 4/Pink=daily consumption\label{t0fig}}
\end{figure}

In spite of having three different outcomes, only two global patterns were identified in the final model (Figure \ref{t0fig}). Sixty-one of these foods had a shared modal consumption across both global patterns. This included shared trends such as daily consumption of coffee and foods with added oils; weekly consumption of chicken, refined grain rice, cheese, eggs, cooked dried beans, tomatoes, mixed vegetables, and bananas; and no consumption of lettuce salads, cereals, pies, and cottage cheese. Between the two global profiles we saw a greater favor for fruits, meats, and snacks in global 1, and a greater favor for vegetables in global profile 2. 

\begin{figure}
\includegraphics[width=0.7\textwidth]{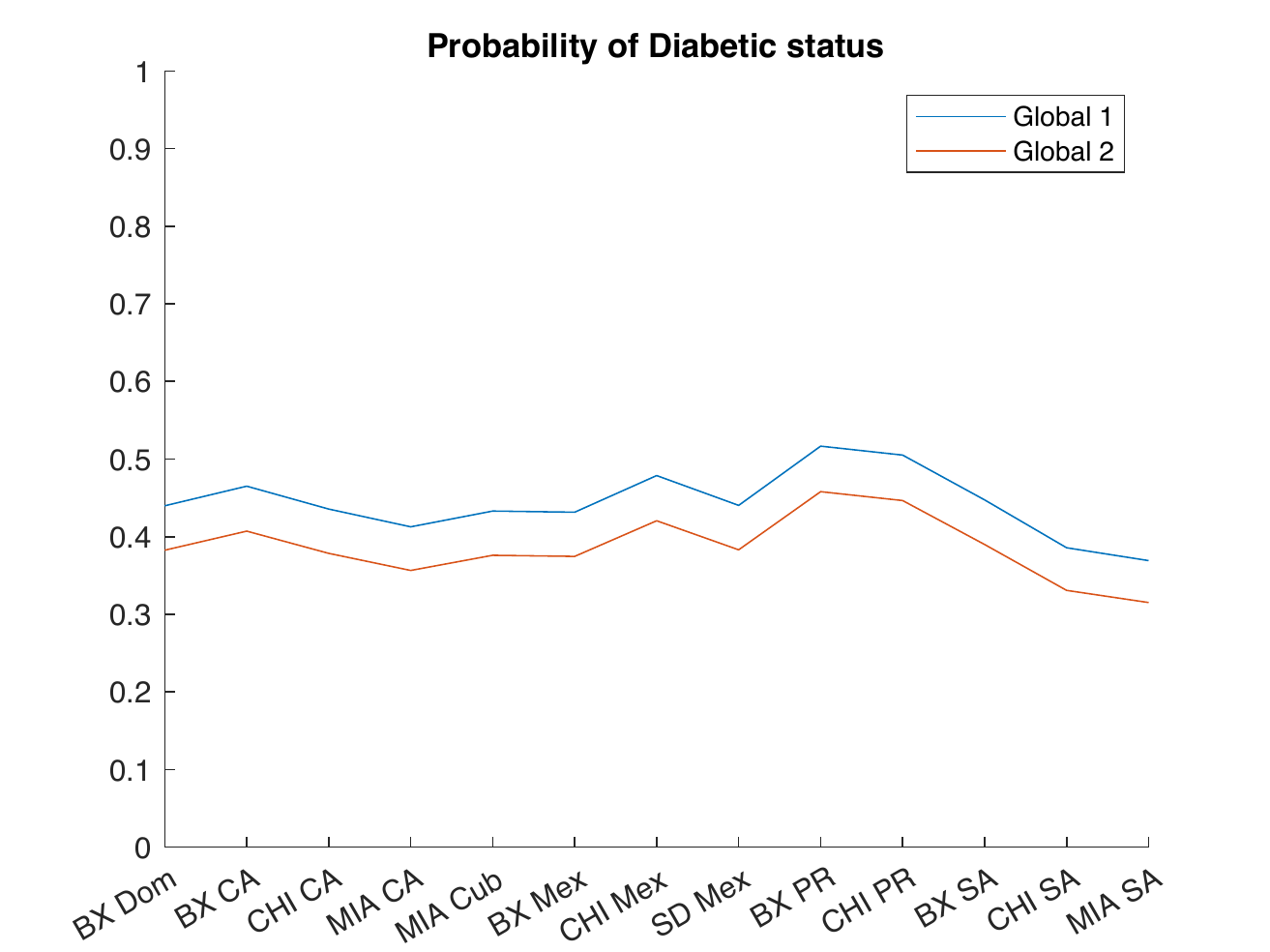}
\caption{Probability of diabetes status for each subpopulation, based on assignment to global diet patterns 1 and 2.\label{probitfig}}
\end{figure}

Global profile 1 was assigned to a majority of participants from 8 of the subpopulations, with participants from Miami (Central American, Cuban, and South American) having at least 70\% of their participants assigned. The remaining subpopulations had a majority of participants assigned to Global profile 2, with 70\% of San Diego participants of Mexican background assigned. Across all subpopulations, participants assigned to global profile 1 had a higher probability of having diabetes and participants assigned to global profile 2 had a higher probability of having a normal or pre-diabetic status. Marginal effects of the two global patterns were both negative with a greater magnitude amongst global 2, indicating a lower probability of more severe diabetes outcomes (e.g. pre-diabetic and diabetic). The probit model boundaries to define those with pre-diabetic status was very small, $\delta_1 =  -1.7\times 10^{-6} (-9.7\times 10^{-5},0.12); \delta_2 = -2.4\times 10^{-6} (-6.5\times 10^{-8},2.0\times10^{-5})$. This indicated little change in effect from for participants with normal and pre-diabetic status, while accounting for global diet patterns. Figure \ref{probitfig} shows how the probability of diabetes status differed by subpopulation. Miami participants of South and Central American background had some of the lowest probabilities, across both global patterns. Participants of Puerto Rican background from Bronx and Chicago had the highest probabilities of diabetes across both global patterns. 

\subsection{Subpopulation-specific dietary patterns}
\begin{figure}
\includegraphics[height=0.9\textheight]{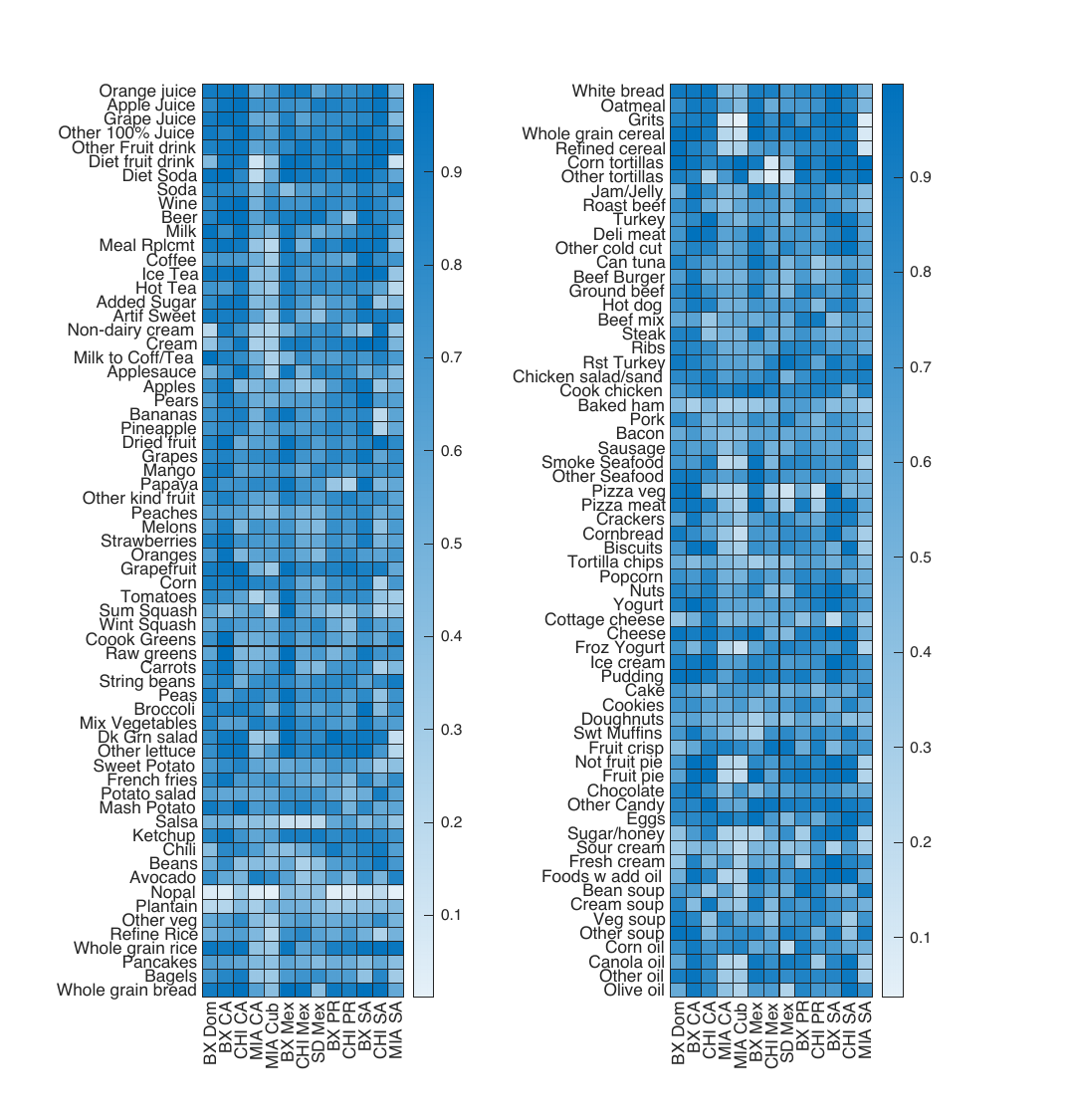}
\caption{Heatmap of global/local allocation indicator, based on a threshold of $\nu_j^{(s)} < 0.4$ is considered local. Foods more likely to be allocated to global level are darker in hue.  \label{hchsfig_nu}}
\end{figure}

While the ordinal supervised RPC model focuses on impact at the global level, we detail differences found at the local level for exploratory purposes. Insight into these local patterns may be useful for further investigation in later research. Local level patterns are identified when subpopulations share a unique departure from the global pattern they were assigned to. Figure \ref{hchsfig_nu} shows the probability of a subpopulation deviating from their global assigned pattern. We select a conservative threshold, focusing on foods that had at least a 60\% probability of deviating from the global pattern. 

Amongst the 13 predefined subpopulations, participants from Miami had the most food consumption patterns that deviated from the overall population ($~64\%$). Bronx participants with Central American background (11\%) and Puerto Rican background (14\%) had the least number of food consumption patterns that deviated from the overall population patterns. Of the 129 foods observed, 53 (41\%) of those foods assumed a pattern at the global level for all 13 subpopulations. All remaining foods had at least one subpopulation which deviated from the overall global patterns. 

We highlight some food patterns which favored a strong deviation from their global patterns defined. Nopal was the only food that favored a local deviation for all subpopulations. Regardless of what global pattern was assigned, participants of Mexican ethnic background favored consuming nopal at least 3-4 times a week. All other HCHS/SOL participants favored no consumption. There were differences in salsa consumption amongst 7 subpopulations. All Miami participants favored no consumption. Bronx participants of Mexican background and Chicago participants of Central American background favored weekly consumption. Participants of Mexican background from Chicago and San Diego  favored daily consumption. Plaintain differed among 6 subpopulations favoring weekly consumption amongst Bronx participants of Dominican, Central American, Puerto Rican, and South American background as well as Miami participants of Central American background. While global patterns favored a monthly and weekly consumption of summer squash, Miami participants of Cuban and South American background, Chicago participants of Puerto Rican and South American background, and Bronx participants of Puerto Rican background all favored no consumption. 

Melons and green peas favored a higher consumption by Chicago participants of South American background (at least 3-4 times per week). Crackers and other kinds of vegetables had a higher consumption by Miami participants of Cuban background (at least 3-4 times per week). Whole grain bread favored a higher consumption by San Diego participants of Mexican background (at least 3-4 times per week). Beef mixtures and steak were consumed more by Chicago participants of Central American background (at least 3-4 times per week). Chili with meat had a higher consumption by Bronx participants of Mexican background (at least 3-4 times per week). Fresh tomatoes favored a higher consumption amongst participants of South American background from Chicago and Miami, as well as Miami participants of Central American background (at least daily).  Tomato/vegetable soup favored weekly consumption by Chicago participants of Central and South American background. Roast beef favored weekly consumption by Miami participants of Central and South American background. Refined grain rice favored daily consumption by Miami participants of Cuban background and Chicago participants of South American background. Sweet potatoes favored weekly consumption by Miami participants of South American background and no consumption by Chicago participants of South American background.

\section{Discussion}
The ordinal supervised robust profile clustering (osRPC) model, discussed in this paper, is an extension of the supervised robust profile clustering model that previously identified shared exposures associated with a binary outcome. The osRPC model permits three or more levels, that increase in severity or level. The model identifies global patterns shared amongst the overall population. The identification and characterization of these patterns is dependent on the ordinal outcome. The flexibility of the RPC allows for the exposure-outcome relationship to focus solely on those exposures that are shared amongst the whole population and remove exposures that are driven locally. This reduction of subpopulation-specific effects from the overall model improved overall pattern identification prediction, as illustrated in our simulation studies. 

Using the Hispanic Community Health Study/Study of Latinos, we applied the osRPC model to identify overall dietary patterns that may be associated with the diabetes status (normal, pre-diabetic, diabetic), accounting for subpopulation differences defined by ethnicity and study site. While removing the noise of any subpopulation-specific patterns that may reduce identifiability for the overall study population, we identified two global dietary patterns. While the two patterns did not show a statistically significant association to the diabetes status, we did see a slight increase in the proportion of participants with diabetes amongst those who were assigned to global pattern 1, which was characterized by more frequent consumption of fruits, snack foods, and breads (pancakes, bagels, etc.). The boundary parameters of the osRPC from normal to pre-diabetic status were very narrow. This small difference between the two levels of the outcome may imply that the two levels could be collapsed to one level. However, we kept the model with three levels, to demonstrate the utility of the model when more than two levels are necessary. Additionally, this case study focused on the association of diet and diabetes, after subpopulation-specific effects had been removed. This relationship was examined using a cell-means coding scheme within the probit regression model to mitigate label-switching amongst the outcome-dependent global dietary patterns, easier interpretation and computational implementation under the probit link function \citep{molitor2010bayesian,stephenson2022derivation}. Additional confounders that may influence this relationship (e.g. age, sex, comorbidities) were not included for this case study, but potentially could be with additional model extensions that rely on the recently introduced mixture reference cell coding scheme with further research \citep{wu2024derivation}. 

Differences in global diet pattern membership were identified across the 13 different subpopulations. Given the dependency on the outcome, it is possible that the differences in membership and association to the outcome could be connected to local dietary behaviors shared within each of the defined subpopulations. With subpopulations defined by both ethnic background and study site, some similarities were found amongst participants who shared ethnic backgrounds (e.g. nopal consumption) or study site (e.g. tomato/vegetable soup consumption in Chicago). However, most patterns were not uniformly shared across all participants of that ethnic background or study site, warranting the need for the cross-section of the two demographic variables to define the subpopulation. Unlike the osLCM, the osRPC focuses on behaviors shared only by the overall population, removing any subpopulation-specific patterns that would otherwise interfere with the primary relationship between the overall diet patterns and the multi-level outcome of interest. Consequently, the association of how these subpopulation-specific diets impact our health outcome is not quantified. The inclusion of the local dietary pattern effects would yield for a very large and unstable regression model. Further research is needed to better understand how to incorporate local effects into the model, while still preserving model stability and computational efficiency. 

We applied our proposed method to the HCHS/SOL study for demonstration without accounting for the complex survey sampling design of the study \cite{Lavange:2010aa}. \cite{wu2024derivation} has implemented a survey-weighted overfitted mixture model that can allow for such integration. However, this approach has not yet been extended to accommodate the flexibility and complexity of the supervised robust profile clustering framework. Such an extension would be a worthwhile future direction. 

A handful of studies have examined dietary pattern differences amongst adult participants of HCHS/SOL \citep{maldonado2021dietary,maldonado2022posteriori,casagrande2018variations,mcclain2018frequency,chen2022healthy,han2020dietary}, but only two have examined those differences by ethnic background and study site \citep{stephenson2020empirically, de2022shared}. \citep{stephenson2020empirically} analyzed nine of our thirteen subpopulations via the unsupervised version of this model (Robust Profile Clustering), excluding Bronx participants of Mexican, Central American and South American background and Chicago participants of South American background. Consistent with our study only two global diet patterns  and one local diet pattern nested within each of the predefined subpopulations were derived.  However, the distribution of these patterns differed greatly from our results due to the lack of dependency on diabetes status. For example, rice consumption favored a deviation to the local level for all 9 subpopulations in \citep{stephenson2020empirically}, with a high frequency consumption for participants from Miami. Our model identified rice as a global leaning food item for all 13 subpopulations favoring a consumption frequency of at least weekly. Papaya consumption deviated to the local level in both studies for participants of Puerto Rican background favoring a likelihood of no consumption, but the global diet patterns differed for all other subpopulations favoring either no consumption (Global profile 2) or at least monthly consumption (Global profile 1). \citep{de2022shared} also analyzed overall and ethnic background-study site defined subpopulation-specific dietary patterns for 12 of our 13 subpopulations, excluding Bronx participants of South American background. The authors also utilized a different dietary instrument (24-hour recalls), different exposure variables (nutrient intake), and a different analytic approach (Multi-Study Factor Analysis). Their analysis also showed deviations from overall shared patterns that differed from our results. For example, seafood intake shared similarities across all subpopulations, with the exception of Bronx participants of Mexican background and Chicago participants of South American background. These two subpopulations favored higher consumption frequencies compared to other subpopulations. This trend was not reflected in our study which favored lower consumptions across all subpopulations, based on our FPQ data queried on intake over the past year as opposed to a summary of 1-2 days of intake. While both dietary assessments rely on self-reporting of the participants, both HCHS/SOL assessments have been validated and  reliable sources of intake \citep{yan2022validity, willett2012nutritional}. However, the inability for 24-hour recalls to capture episodically consumed foods exacerbates the issue of measurement error in capturing complete dietary consumption habits compared to our results reported via the FPQ \citep{yuan2017validity}. Additional information from biomarkers can greatly mitigate the issue of measurement error due to misreporting, but there is substantial
under-reporting of  self-reported energy and protein in HCHS/SOL \citep{mossavar2015applying}.

Mixture model approaches are data dependent. Furthermore, supervised mixture models are dependent on the outcome of interest, which can drive the patterns being characterized. This is clearly illustrated in \citep{stephenson2020empirically} where the same dietary exposure was analyzed accounting for similar subpopulations but independent of any outcome. \citep{corsino2017association} assessed the diet-diabetes relationship in HCHS/SOL participants. However, intake was assessed by a participants adherence to a pre-defined recommended dietary pattern, the Dietary Approach to Stop Hypertension (DASH) using data from 24-hour recall data. Participants with greater adherence to the DASH score had a high consumption of fruits, vegetables, and whole grains and low consumption of meats and were associated with lower insulin resistance, lowering diabetes risk. While not data-driven like our approach, we found the results consistent to what was highlighted in our global patterns. Global profile 2 with a lower proportion of diabetic individuals favored more frequent consumptions of fruits and vegetables and lower frequent consumptions of meats compared to global profile 1. 

The use of the DASH score in \citep{corsino2017association} limits the understanding of diet heterogeneity within the study population and focuses instead on key foods that should be consumed in order to reduce adverse health outcomes of interest. Further, this score is then analyzed alongside the diabetes status to understand how adherence to those food recommendations is associated with diabetes status. \citep{casagrande2018variations} focused on intake of key macro- and micronutrients as reported on 24-hour recalls and its association with glycemic status in HCHS/SOL participants.

Our data-driven approach differs significantly from prior work, allowing for the heterogeneity of multivariate exposures (e.g. diet) within a population to be explored and contextualized jointly with the knowledge of an ordinal level outcome. Our extension allowing the introduction of a multi-level ordinal outcome provides greater insight into how patterns derived from a wide set of exposures (e.g. diet) can change based on the strength of the association and increasing severity of a health outcome.  While much research is still to be pursued in this area, our model adds to the literature of flexible model-based clustering to better understand a large set of exposures and its association on key outcomes of interest. 

%
%

\section*{Data Availability and Code}
Data utilized from the Hispanic Community Health Study/Study of Latinos  are publicly available through a Data and Materials Distribution Agreement (DMDA). To obtain access, applications can be requested and submitted directly from the \href{https://sites.cscc.unc.edu/hchs/}{study website}. Supporting code and data to implement the methods performed in this study are made available on a public repository at \url{https://github.com/bjks10/osRPC/}.

\begin{acks}[Acknowledgments]
The authors thank the staff and participants of HCHS/SOL for their important contributions.
\end{acks}
\begin{funding}
 The Hispanic Community Health Study/Study of Latinos is a collaborative study supported by contracts from the National Heart, Lung, and Blood Institute (NHLBI) to the University of North Carolina (HHSN268201300001I / N01-HC-65233), University of Miami (HHSN268201300004I / N01-HC-65234), Albert Einstein College of Medicine (HHSN268201300002I / N01-HC-65235), University of Illinois at Chicago (HHSN268201300003I / N01- HC-65236 Northwestern University), and San Diego State University (HHSN268201300005I / N01-HC-65237). The following Institutes/Centers/Offices have contributed to the HCHS/SOL through a transfer of funds to the NHLBI: National Institute on Minority Health and Health Disparities, National Institute on Deafness and Other Communication Disorders, National Institute of Dental and Craniofacial Research, National Institute of Diabetes and Digestive and Kidney Diseases, National Institute of Neurological Disorders and Stroke, NIH Institution-Office of Dietary Supplements.
\end{funding}

\newpage
\begin{supplement}
\stitle{Supplementary Figure 1: Distribution of FPQ food items included for analysis}
\begin{figure*}[h]
\begin{subfigure}[h]{0.4\linewidth}
\includegraphics[width=\linewidth]{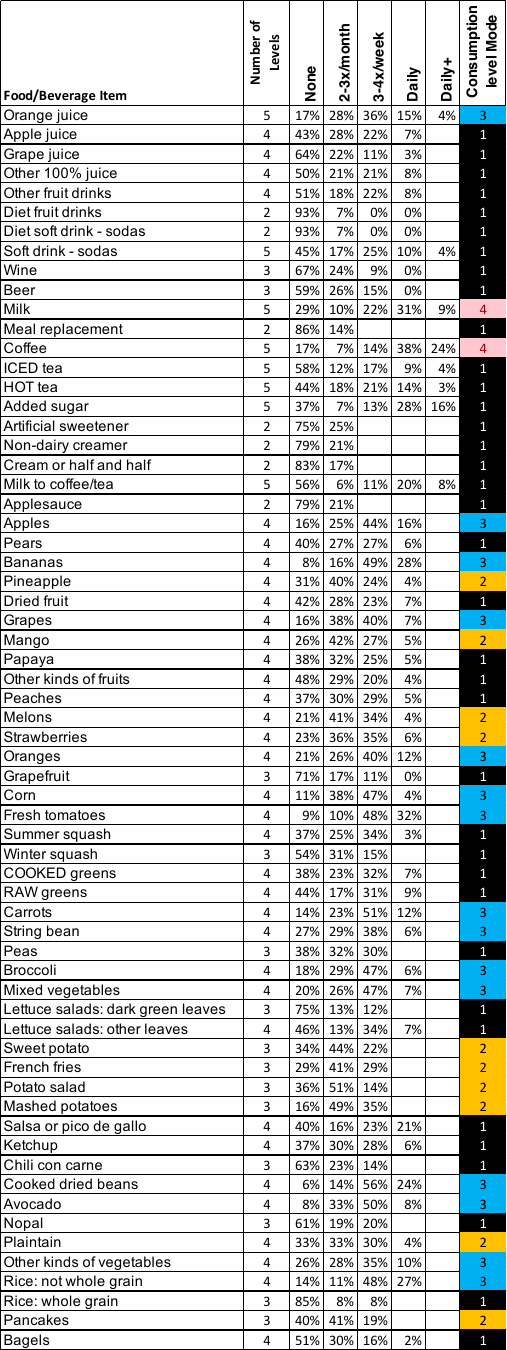}
\end{subfigure}
\hfill
\begin{subfigure}[h]{0.4\linewidth}
\includegraphics[width=\linewidth]{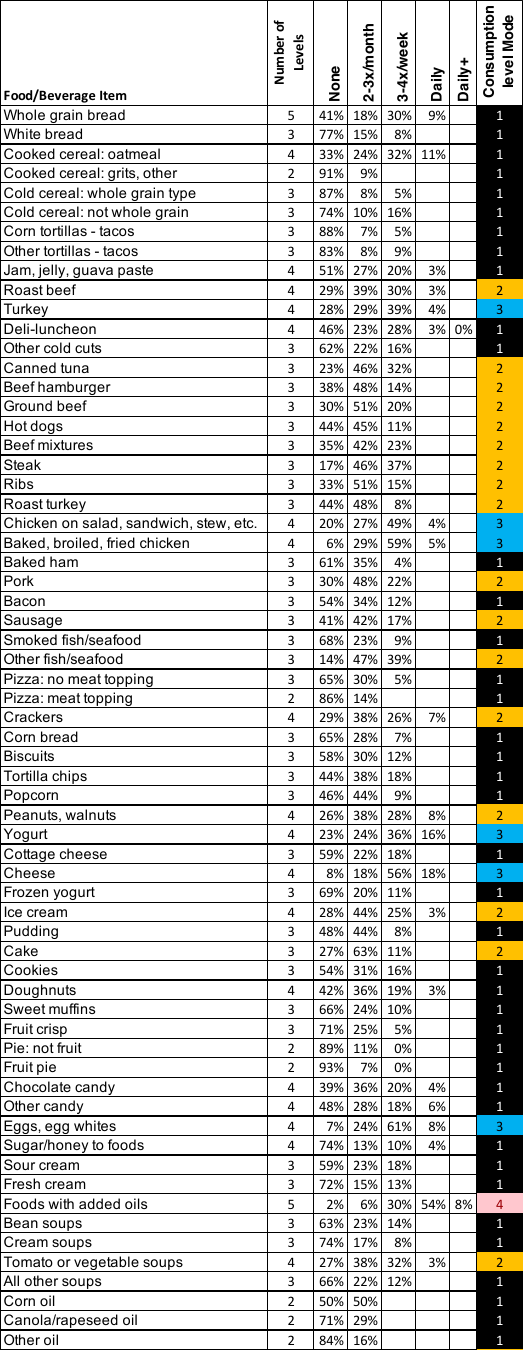}
\end{subfigure}%
\caption{Observed frequency distribution of collapsed 129 derived food items from Food Propensity Questionnaire for 15,504 HCHS/SOL participants. Last column provides the consumption level with the highest proportion level amongst the participants (consumption mode)}
\end{figure*}

\newpage
\stitle{Supplementary Table 1: Demographic Characteristics of HCHS/SOL cohort analyzed}
\begin{table}
\caption{Demographic characteristics of HCHS/SOL cohort included for analysis. Cells contain marginal percentages of cohort}\label{dem-t}
	\begin{tabular}{@{}| l | l |c| c | c| c|@{}}
	\hline
	\multicolumn{2}{l}{Demographic Characteristic} & Overall & Normal &Pre-Diabetic & Diabetic \\
	\multicolumn{2}{l}{}  & N = 12032 & N=4838  & N=4634 & N=2560  \\
	\multicolumn{2}{l}{}  & \% & \% & \% & \% \\
	\hline 
	\multirow{2}{*}{Sex}  	& Male & 38.6 &  37.0 &  40.9 & 37.5 \\
					& Female &  61.4&  63.0 & 59.1 & 62.5 \\
	\hline
	\multirow{4}{*}{Employment} & Retired & 9.8 & 3.8 &  10.1 & 20.7 \\
						& Unemployed & 39.5 & 41.2 & 36.1 & 42.6 \\	
						& Part-Time & 16.9 & 19.5 & 16.9 & 12.0 \\
						& Full-Time & 33.7 & 35.5 & 36.9 & 24.7\\
	\hline
	\multirow{3}{*}{Education} & $<$ High School/GED & 37.7 &  29.6 & 39.9 & 49.2 \\
						& High School/GED & 25.3 & 28.5 & 24.4 & 20.7 \\
					& $\ge$ High School/GED & 37.0& 41.9 & 35.7 & 30.1 \\
	\hline
	\multirow{3}{*}{Alcohol Use}	& Never & 53.5 & 48.7 & 52.3 & 64.6 \\
						& Former & 41.9 & 46.2 & 43.0 & 32.0 \\
						& Current & 4.6 & 5.2 & 4.8 & 3.4\\
	\hline
	\multirow{3}{*}{Income} 	& less than \$30,000 & 64.8 & 62.6 &  64.6 & 69.2 \\
						& at least \$30,000 & 29.9 &  31.9 & 30.9 & 24.3 \\
						& Missing & 5.3 & 5.5 & 4.5 & 6.6 \\
	\hline
	\multirow{2}{*}{Depressed} 	& No & 71.2 & 73.7 & 72.1 & 64.7 \\
						& Yes & 28.8 & 26.3 & 27.9 & 35.3 \\
	\hline
	\multirow{2}{*}{Age} 		& less than 45 years & 38.4 & 60.3 & 29.3 & 13.4 \\
						& at least 45 years & 61.6  & 39.7 & 70.7 & 86.6  \\
	\hline
	\multirow{3}{*}{Nativity} 	& US Born & 15.3 & 20.5 & 12.8 & 9.9 \\
						& $<$ 10 yrs in US & 23.7 & 29.9 & 22.1 & 14.8 \\
						& $\ge$ 10 yrs in US & 61.0 & 49.6 & 65.1 & 75.3 \\
	\hline
	\multirow{2}{*}{Obesity}		& No & 57.3 & 69.6 & 53.0 & 41.8\\
						& Yes & 42.7 & 30.4 & 47.0 & 58.2 \\
	\hline
	\multirow{13}{*}{Subpopulation} & Dominican (BX) & 9.0 & 9.4 & 8.8 & 8.4 \\
						& C. American (BX) & 1.4 & 1.3 & 1.4 & 1.6 \\	
						& C. American (CHI) & 2.5 & 2.5 & 2.6 & 2.3 \\
						& C. American (MIA) & 6.9 & 7.6 & 7.1 & 5.5 \\
						& Cuban (MIA) & 16.1 & 16.4 &17.0 & 13.8 \\
						& Mexican (BX) & 1.3 & 1.2 & 1.6 & 0.9 \\
						& Mexican (CHI) & 14.8 & 13.7 & 14.7 &  17.1 \\
						& Mexican (SD) & 25.0 & 26.4 & 24.3 & 23.5 \\
						& Puerto Rican (BX) & 11.5 & 9.5 & 10.9 & 16.2\\
						& Puerto Rican (CHI) & 5.0 & 4.2 & 5.0 & 6.4 \\
						& S. American (BX) & 1.2& 1.2 &  1.2 & 1.2 \\
						& S. American (CHI) & 2.1 & 2.5 & 2.1 & 1.3 \\
						& S. American (MIA) & 3.3 &  4.2 & 3.3 & 1.8 \\	
	\hline	
	\end{tabular}
\end{table}

\end{supplement}
\newpage

\bibliographystyle{imsart-nameyear} 
\bibliography{ms650refs.bib}       


\end{document}